\documentclass{article} %
\usepackage{iclr2026_conference,times}

\usepackage[utf8]{inputenc} %
\usepackage[T1]{fontenc}    %
\usepackage{hyperref}       %
\usepackage{url}            %
\usepackage{booktabs}       %
\usepackage{amsfonts}       %
\usepackage{nicefrac}       %
\usepackage{microtype}      %
\usepackage{xcolor}  
\usepackage{enumitem}

\usepackage[utf8]{inputenc} %

\usepackage{graphicx}

\usepackage[T1]{fontenc}

\IfFileExists{headers/config/showoverfull.config}{
	\overfullrule=1cm
}{
}

\usepackage{marginnote}

\usepackage[backgroundcolor=none,linecolor=red,textsize=footnotesize]{todonotes}

\usepackage{etoolbox}

\newbool{includeappendix}
\setbool{includeappendix}{true} %
\IfFileExists{headers/config/noappendix.config}{
	\setbool{includeappendix}{false}
}{}

\newif\ifincludeappendixx
\ifbool{includeappendix}{
	\includeappendixxtrue
}{
	\includeappendixxfalse
}

\usepackage{xr} %
\usepackage{filecontents}

\ifbool{includeappendix}{}{
	\input{appendix-labels-loader}

	\externaldocument{appendix-labels}
}

\usepackage{xcolor} %

\definecolor{my-full-blue}{HTML}{1F77B4}

\definecolor{my-full-orange}{HTML}{FF7F0E}

\definecolor{my-full-green}{HTML}{2CA02C}

\definecolor{my-full-red}{HTML}{d62728}

\definecolor{my-full-purple}{HTML}{9467bd}

\definecolor{my-full-brown}{HTML}{8c564b}

\definecolor{my-full-pink}{HTML}{e377c2}

\definecolor{my-full-gray}{HTML}{7f7f7f}

\definecolor{my-full-olive}{HTML}{bcbd22}

\definecolor{my-full-cyan}{HTML}{17becf}

\colorlet{my-blue}{my-full-blue!30}
\colorlet{my-orange}{my-full-orange!30}
\colorlet{my-green}{my-full-green!30}
\colorlet{my-red}{my-full-red!30}
\colorlet{my-purple}{my-full-purple!30}
\colorlet{my-brown}{my-full-brown!30}
\colorlet{my-pink}{my-full-pink!30}
\colorlet{my-gray}{my-full-gray!30}
\colorlet{my-olive}{my-full-olive!30}
\colorlet{my-cyan}{my-full-cyan!30}

\definecolor{my-navy-blue}{HTML}{006EB8}

\usepackage{listings}

\usepackage{textcomp}

\usepackage{xcolor}

\usepackage[scaled=0.8]{beramono}

\definecolor{ckeyword}{HTML}{7F0055}
\definecolor{ccomment}{HTML}{3F7F5F}
\definecolor{cstring}{HTML}{2A0099}

\lstdefinestyle{numbers}{
	numbers=left,
	framexleftmargin=20pt,
	numberstyle=\tiny,
	firstnumber=auto,
	numbersep=1em,
	xleftmargin=2em
}

\lstdefinestyle{layout}{
	frame=none,
	captionpos=b,
}

\lstdefinestyle{comment-style}{
	morecomment=[l]//,
	morecomment=[s]{/*}{*/},
	commentstyle={\color{ccomment}\itshape},
}

\lstdefinestyle{string-style}{
	morestring=[b]",%
	morestring=[b]',%
	stringstyle={\color{cstring}},
	showstringspaces=false,%
}

\lstdefinestyle{keyword-style}{
	keywordstyle={\ttfamily\bfseries},
	morekeywords={
		function,
		constructor,
		int,
		bool,
		return,
		returns,
		uint
	},
	morekeywords = [2]{},
	keywordstyle = [2]{\text},
	sensitive=true,
}

\lstdefinestyle{input-encoding}{
	inputencoding=utf8,
	extendedchars=true,
	literate=
	{ℝ}{$\reals$}1%
	{→}{$\rightarrow$}1%
	{α}{$\alpha$}1%
	{β}{$\beta$}1%
	{λ}{$\lambda$}1%
	{θ}{$\theta$}1%
	{ϕ}{$\phi$}1%
}

\lstdefinestyle{escaping}{
	moredelim={**[is][\color{blue}]{\%}{\%}},
	escapechar=|,
	mathescape=true
}

\lstdefinestyle{default-style}{
	basicstyle=\fontencoding{T1}\ttfamily\footnotesize,
	style=numbers,
	style=layout,
	style=comment-style,
	style=string-style,
	style=keyword-style,
	style=input-encoding,
	style=escaping,
	tabsize=2,
	upquote=true
}

\lstdefinelanguage{BASIC}{
	language=C++,
	style=default-style
}[keywords,comments,strings]%

\usepackage{microtype}
\usepackage{graphicx}
\usepackage{booktabs}
\usepackage{threeparttable}
\usepackage{adjustbox}
\usepackage{multirow}
\usepackage{xspace}
\usepackage{wrapfig}
\usepackage{makecell}
\usepackage{dsfont}
\usepackage{subcaption}

\usepackage{hyperref}

\usepackage{amsmath,amsfonts,bm}
\usepackage{braket}

\def\eqref#1{equation~\ref{#1}}

\def\1{\bm{1}}

\def\vzero{{\bm{0}}}

\def\vtheta{{\bm{\theta}}}
\def\vTheta{{\bm{\Theta}}}
\def\vphi{{\bm{\phi}}}

\def\vdelta{{\bm{\delta}}}

\def\vb{{\bm{b}}}

\def\vf{{\bm{f}}}

\def\vx{{\bm{x}}}

\DeclareMathAlphabet{\mathsfit}{\encodingdefault}{\sfdefault}{m}{sl}
\SetMathAlphabet{\mathsfit}{bold}{\encodingdefault}{\sfdefault}{bx}{n}

\usepackage{amsmath}
\usepackage{amssymb}
\usepackage{mathtools}
\usepackage{amsthm,thmtools}
\usepackage{fontawesome5}

\theoremstyle{plain}

\theoremstyle{definition}

\theoremstyle{remark}

\usepackage{tikz}
\usetikzlibrary{calc,decorations,decorations.pathmorphing}
\usetikzlibrary{positioning,fit,arrows}
\usetikzlibrary{decorations.markings}
\usetikzlibrary{shapes,shapes.geometric}
\usetikzlibrary{shadows,patterns,snakes}
\usetikzlibrary{backgrounds,decorations.pathreplacing,automata}
\usetikzlibrary{intersections}
\usetikzlibrary{angles,quotes}
\usetikzlibrary{plotmarks}
\usetikzlibrary{patterns}
\usetikzlibrary{arrows.meta}
\usetikzlibrary{shapes.misc}
\usetikzlibrary{chains}
\usepackage{siunitx}
\newcolumntype{d}[1]{S[table-format=#1]}
\usetikzlibrary{patterns.meta}

\usepackage[capitalize,noabbrev]{cleveref}

\makeatletter
\newcommand\theHALG@line{\thealgorithm.\arabic{ALG@line}}
\makeatother

\crefformat{section}{\S#2#1#3}

\crefrangeformat{section}{\S#3#1#4\crefrangeconjunction\S#5#2#6}

\crefmultiformat{section}{\S#2#1#3}{\crefpairconjunction\S#2#1#3}{\crefmiddleconjunction\S#2#1#3}{\creflastconjunction\S#2#1#3}

\newcommand{\crefrangeconjunction}{--}

\crefname{listing}{Lst.}{listings}
\crefname{line}{Lin.}{Lin.}
\crefname{appendix}{App.}{App.}

\newcommand{\appref}[1]{%
	\ifbool{includeappendix}{\cref{#1}}{the appendix}%
}
\newcommand{\Appref}[1]{%
	\ifbool{includeappendix}{\cref{#1}}{The appendix}%
}

\title{Practical Hybrid Quantum Language Models with Observable Readout on Real Hardware}

\author{Stefan Balauca$^{1,2}$ \quad Ada-Astrid Balauca$^{2}$ \quad Adrian Iftene$^{1}$\\
$^{1}$ Department of Computer Science, ``Al. I. Cuza'' University, Iasi, Romania\\
$^{2}$ FreeYaMind Campus, Iasi, Romania\\
\texttt{\{stefan-razvan.balauca,adiftene\}@info.uaic.ro} \\
}

\AddToHook{cmd/appendix/before}{%
    \crefalias{section}{appendix}%
    \crefalias{subsection}{appendix}
}

\iclrpreprintcopy %
\begin{document}

\maketitle

\begin{abstract}
Hybrid quantum-classical models represent a crucial step toward leveraging near-term quantum devices for sequential data processing.
We present Quantum Recurrent Neural Networks (QRNNs) and Quantum Convolutional Neural Networks (QCNNs) as hybrid quantum language models, reporting the first empirical demonstration of generative language modeling trained and evaluated end-to-end on real quantum hardware.
Our architecture combines hardware-optimized parametric quantum circuits with a lightweight classical projection layer, utilizing a multi-sample SPSA strategy to efficiently train quantum parameters despite hardware noise.
To characterize the capabilities of these models, we introduce a synthetic dataset designed to isolate syntactic dependencies in a controlled, low-resource environment.
Experiments on IBM Quantum processors reveal the critical trade-offs between circuit depth and trainability, demonstrating that while noise remains a significant factor, observable-based readout enables the successful learning of sequential patterns on NISQ devices.
These results establish a rigorous engineering baseline for generative quantum natural language processing, validating the feasibility of training complex sequence models on current quantum hardware.
\end{abstract}

\section{Introduction}

Natural language processing (NLP) has seen remarkable advances in recent years, largely driven by deep learning architectures such as recurrent neural networks (RNNs) \citep{elman1990finding,hochreiter1997long}, convolutional neural networks (CNNs) \citep{kim2014convolutional}, and Transformers \citep{vaswani2017attention}. These models have enabled impressive performance across a range of tasks, including language modeling, machine translation, and text generation. However, the computational cost of scaling these architectures has motivated the exploration of alternative paradigms that might offer different inductive biases or representational efficiencies.

Quantum computing has emerged as a candidate to broaden the scope of machine learning by leveraging the principles of superposition and entanglement \citep{nielsen2010quantum,schuld2015introduction}. In particular, hybrid quantum-classical models, where parametric quantum circuits (PQCs) are combined with classical post-processing, have attracted significant attention due to their compatibility with noisy intermediate-scale quantum (NISQ) devices \citep{preskill2018quantum,benedetti2019parameterized}. Theoretically, these models offer access to exponentially large Hilbert spaces for feature representation; experimentally, however, they remain constrained by device noise, limited connectivity, and the difficulty of training deep circuits.

While a variety of quantum approaches to NLP have been proposed, most works remain largely theoretical or validate methods only on noiseless simulators. The DisCoCat framework \citep{coecke2010mathematical} pioneered a categorical connection between semantics and quantum circuits, leading to experimental demonstrations of sentence classification on real hardware \citep{lorenz2023qnlp}. However, classification tasks compress a sequence into a single label, bypassing the complexities of sequential generation. Systematic investigations of quantum sequence models for \textit{generative} tasks, which require the model to maintain coherent memory of syntactic structure over time, remain scarce and have not yet been demonstrated on physical quantum processors.

Our work addresses this gap by presenting the first end-to-end training and evaluation of Quantum Recurrent Neural Networks (QRNNs) and Quantum Convolutional Neural Networks (QCNNs) for generative language modeling on NISQ devices. Moving beyond proof-of-principle simulations, we design hardware-optimized circuits adapted to the heavy-hex topology of IBM quantum processors. Our hybrid quantum language models (HQLMs) combine these parametric circuits with a lightweight classical projection layer, utilizing a hardware-friendly multi-sample SPSA strategy to estimate gradients through the quantum noise. We evaluate these architectures on both simulators and real hardware, providing a rigorous analysis of the trade-offs between circuit depth, expressivity, and hardware noise in the context of next-word prediction.

\vspace{-3mm}
\paragraph{Contributions} Our work makes the following contributions:
\vspace{-2mm}
\begin{enumerate}[leftmargin=*, itemsep=0pt, topsep=0pt]
    \item We propose \textbf{hybrid quantum language models} based on QRNNs and QCNNs, specifically adapted for generative sequence modeling on heavy-hex quantum hardware.
    \item We introduce a \textbf{scalable training framework} using multi-sample SPSA for quantum parameters and gradient-based updates for classical layers, enabling end-to-end training on noisy devices.
    \item We analyze the role of \textbf{quantum embeddings} and observable-based feature extraction (Z and ZZ), quantifying architectural trade-offs between circuit depth, qubit count, and trainability.
    \item We provide an \textbf{empirical evaluation} on synthetic natural language datasets, establishing a performance baseline for generative quantum models and characterizing their robustness to physical noise compared to classical simulation.
    \item We report the first \textbf{experimental demonstration} of training hybrid sequence models on real quantum hardware for generative language modeling, validating the practical feasibility of these architectures in the NISQ era.
\end{enumerate}

\begin{figure}[t]
    \centering
    \includegraphics[width=0.9\textwidth]{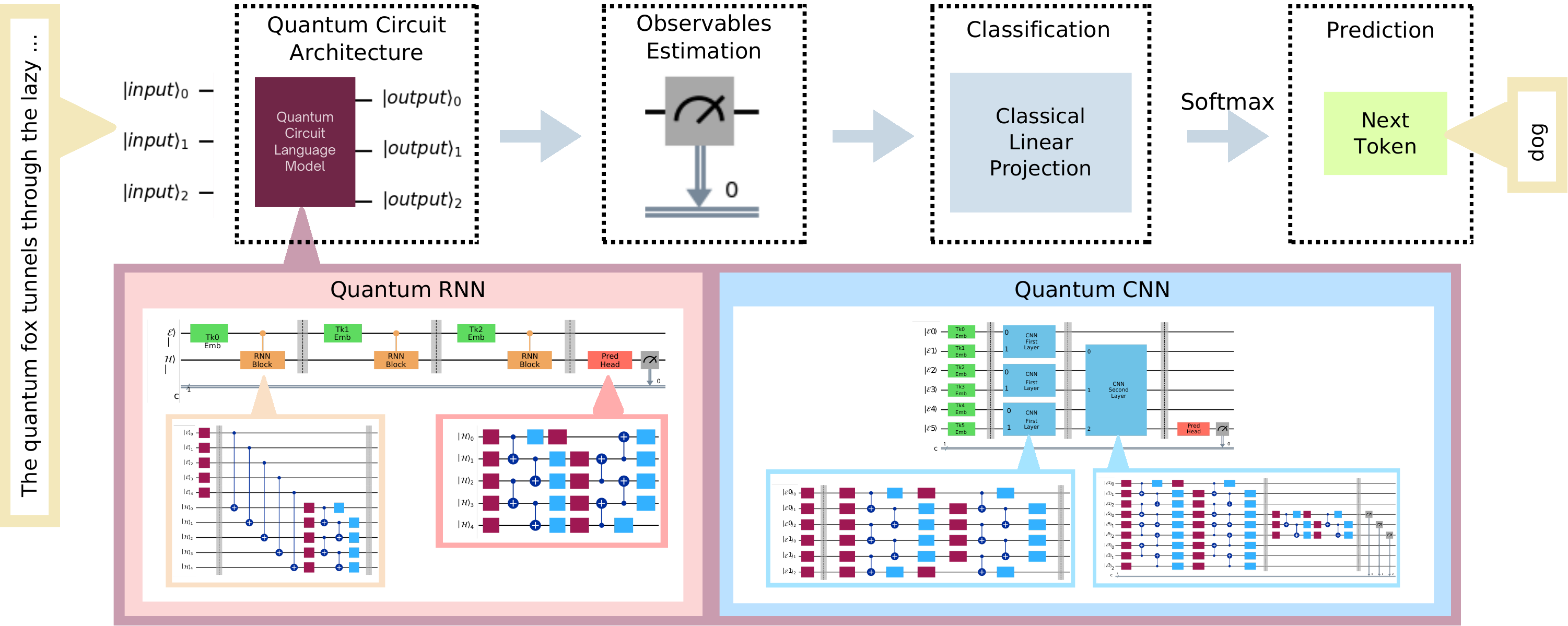}
    \caption{Overview of our hybrid quantum language models (HQLMs). Tokens are embedded into quantum states, processed by QRNN or QCNN layers, and mapped to predictions via a classical projection head, trained end-to-end with multi-sample SPSA and gradient-based updates.}
    \label{fig:architecture_overview}
    \vspace{-4mm}
\end{figure}

Our results suggest that while current hardware noise remains a bottleneck, hybrid quantum architectures can successfully learn sequential dependencies, offering a tangible foundation for exploring quantum advantages in NLP as hardware fidelity improves.

\vspace{-2mm}
\section{Background and Notation}
\vspace{-1mm}
\label{sec:background}
We briefly introduce the concepts underlying our hybrid quantum language models (HQLMs).  
Notation details are provided in \Cref{app:notation}.

\vspace{-1mm}
\subsection{Language Modeling}
\label{subsec:lm}
\vspace{-1mm}

Language modeling estimates the probability distribution of token sequences.  
Given a sequence of tokens $(x_1, \dots, x_T)$, a model learns to predict the probability
\vspace{-1mm}
\begin{equation}
    P(x_1, \dots, x_T) = \prod_{t=1}^T P(x_t \mid x_{<t}),\vspace{-1mm}
\end{equation}
typically in an autoregressive fashion.  
The objective is to capture syntactic and semantic dependencies so that the model can predict the next token from context.  
Evaluation metrics such as cross-entropy loss or perplexity measure alignment with observed sequences.  

\vspace{-1mm}
\subsection{Classical Neural Architectures}
\label{sec:classical-nn-lm}
\vspace{-1mm}

Classical neural language models map token embeddings to next-token probabilities using architectures that model sequential structure.  
\emph{Recurrent networks} (RNNs, LSTMs, GRUs) process tokens step-by-step via hidden states.  
\emph{Convolutional networks} (CNNs) apply 1D convolutions to capture local n-gram patterns, extended with dilations for longer contexts.  
\emph{Transformers} replace recurrence with attention, modeling pairwise dependencies across the sequence.  

\vspace{-1mm}
\subsection{Quantum Computing Basics}
\label{sec:qc-intro}
\begin{figure}[t]
  \centering
  \begin{minipage}[c]{0.45\textwidth}
    \centering
    \vspace{3.5mm}
    {\includegraphics[trim={11mm 0 2mm 0},clip,width=1.0\linewidth]{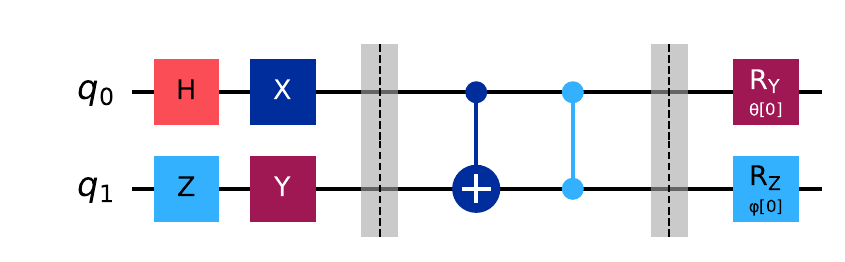}}
    \caption{Examples of elementary quantum gates, the building blocks of quantum circuits: single-qubit gates ($H$, $X$, $Y$, $Z$), two-qubit entangling gates (CNOT, CZ), and single-qubit parametric gates ($R_Y(\theta)$, $R_Z(\phi)$).}
    \label{fig:basic_gates}
  \end{minipage}\hfill
  \begin{minipage}[c]{0.53\textwidth}
    \centering
    \includegraphics[width=0.95\linewidth]{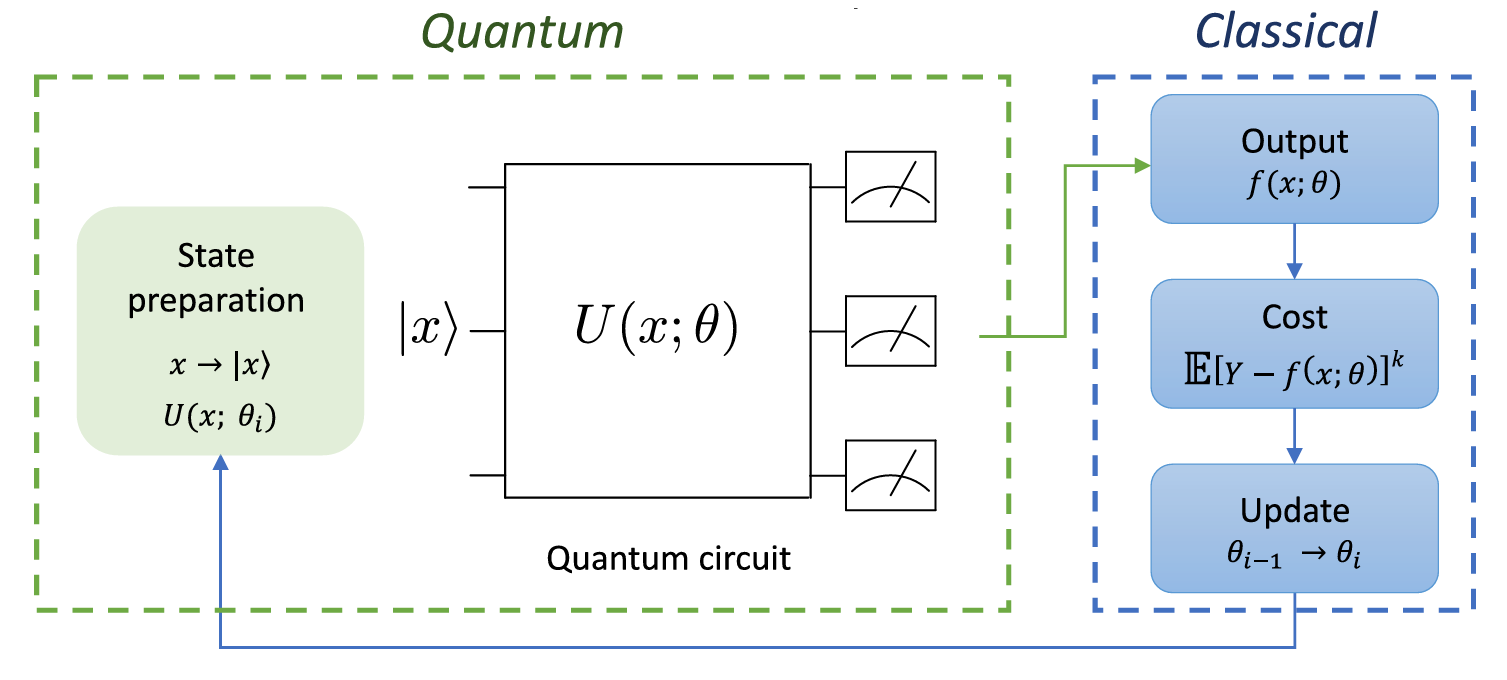}
    \vspace{-3mm}
    \caption{Schematic of a Variational Quantum Algorithm (VQA) workflow from \citet{macaluso2020variational}: a parametric quantum circuit (PQC) is optimized by a classical iterative algorithm.}
    \label{fig:vqa_overview}
  \end{minipage}
  \vspace{-4mm}
\end{figure}

\vspace{-1mm}
Quantum computation is based on \emph{qubits}, which can exist in superpositions such as
\vspace{-0.5mm}
\begin{equation}
\ket{\psi} = \alpha\ket{0} + \beta\ket{1}, \quad |\alpha|^2+|\beta|^2=1.
\vspace{-1mm}
\end{equation}
Multiple qubits form states in tensor-product spaces, enabling \emph{entanglement}, i.e., correlations without classical analogues.  
Quantum gates are unitary operators; standard examples include the Hadamard ($H$), Pauli-$X$, and the entangling CNOT.  
\cref{fig:basic_gates} shows the elementary building blocks, including parametrized single-qubit rotations $R_Y(\theta)$ and $R_Z(\phi)$.

\vspace{-1mm}
\subsection{Parameterized Quantum Circuits (PQCs)}
\label{subsec:pqc}
\vspace{-1mm}
Parameterized quantum circuits (PQCs), also known as variational circuits, are the quantum analogue of neural networks.  
They consist of input encodings, layers of parametrized gates $U(\vtheta)$, and measurement.  
Given a classical input $\vx$, the circuit produces expectation values
\vspace{-0.5mm}
\begin{equation}
f_\vtheta(\vx) = \bra{\phi(\vx)} U^\dagger(\vtheta) O U(\vtheta) \ket{\phi(\vx)},\vspace{-0.5mm}
\end{equation}
where $O$ is an observable (e.g., $Z$ or $ZZ$ operators).  
The parameters $\vtheta$ are trained in a hybrid loop with a classical optimizer.  
An overview of this framework is illustrated in \Cref{fig:vqa_overview}.

\vspace{-1mm}
\subsection{Readout on NISQ Devices}
\label{subsec:nisq}
\vspace{-1mm}

Noisy intermediate-scale quantum (NISQ) devices are limited by circuit depth and gate errors, making PQC design a trade-off between expressivity and noise resilience.  
A key choice is how to extract classical features from the quantum state.  

The \emph{sampling approach} measures computational basis states directly, interpreting outcomes as token probabilities.  
While intuitive, it often complicates optimization, since probability mass must align with discrete encodings.  

The alternative, used here, is \emph{estimator-based readout}: computing expectation values of observables to obtain continuous feature vectors.  
These features are then mapped by a classical linear layer into token logits: $\ell = W \vf + \vb$, with $W \in \mathbb{R}^{d \times V}$.  
This hybrid readout smooths optimization and integrates naturally with classical training while remaining feasible on real hardware.

\vspace{-1mm}
\section{Related Work}
\vspace{-1mm}
\subsection{Quantum Machine Learning (QML)}
\vspace{-1mm}
Quantum machine learning studies how quantum circuits can augment or replace elements of classical learning algorithms \citep{biamonte2017qml,schuld2015introduction}.  
A central paradigm is the use of variational quantum circuits (VQCs), where parametrized gates define expressive models trained with classical optimization \citep{mcclean2016theory,schuld2020circuit,benedetti2019parameterized}.  
These hybrid quantum-classical approaches are well-suited for noisy intermediate-scale quantum (NISQ) devices \citep{preskill2018quantum}, as they exploit quantum superposition and entanglement while relying on lightweight classical layers for readout and stability.  

VQCs face distinctive challenges, including barren plateaus \citep{mcclean2018barren}, which motivate research into circuit expressibility and entangling capacity \citep{sim2019expressibility}, as well as mitigation strategies such as architectural constraints or residual connections \citep{kashif2024resqnets}.  
Beyond classification and regression, quantum circuits have been extended to recurrent models \citep{bausch2020recurrent,macaluso2020variational}, convolutional designs \citep{cong2019quantum}, and natural language tasks \citep{coecke2010mathematical,meichanetzidis2020quantum,lorenz2023qnlp}.  
Surveys provide broader overviews of algorithms and applications \citep{jerbi2023quantum,chen2024survey,nausheen2025quantum}.  
Nevertheless, most work remains simulator-based or limited to small-scale proofs of concept, leaving open the question of whether hybrid quantum models can be trained end-to-end on real hardware.  

\vspace{-1mm}
\subsection{Quantum Natural Language Processing (QNLP)}
\vspace{-1mm}
Quantum NLP builds on the compositional distributional (DisCoCat) framework \citep{coecke2010mathematical}, which maps grammatical structure to tensor networks and, in turn, quantum circuits. Early works explored variational circuits for question answering and entailment \citep{lorenz2023qnlp, meichanetzidis2020quantum}, quantum embeddings for token representations \citep{panahi2019word2ket, chen2021quantum}, and bag-of-words style models \citep{lorenz2023qnlp}. Notably, \citet{karamlou2022quantum} approached quantum natural language generation via combinatorial optimization. While these approaches demonstrate feasibility, they are typically limited to shallow circuits, small-scale tasks, or simulators, and do not address full sequence modeling.

In parallel, recent studies investigate quantum adaptations of Transformer architectures, including residual designs and attention mechanisms \citep{khatri2024quixer, liao2024gpt, amire2025quantumgptmini, tomal2025quantum}. Although promising, these models often require deeper circuits and more qubits, making them challenging to deploy on current NISQ devices. In contrast, we focus on QRNNs and QCNNs, which provide a more hardware-efficient approach to sequential quantum processing. Our work complements prior efforts by presenting the first end-to-end training and evaluation of hybrid quantum sequence models for generative language modeling on real quantum hardware.

\vspace{-1mm}
\subsection{Quantum Recurrent Neural Networks (QRNNs)}
\vspace{-1mm}
QRNNs implement sequential processing through repeated application of parametrized unitaries acting on two registers: a short-lived \emph{embedding} register (encoding the current token) and a longer-lived \emph{hidden} register carrying memory across time steps \citep{bausch2020recurrent,meichanetzidis2020quantum}.  
Each step prepares an embedding state, entangles it with the hidden register via a recurrent block, and updates the hidden state for the next step.  
Optimization typically combines gradient-free methods or parameter-shift rules with classical output layers \citep{schuld2020circuit,jerbi2023quantum}.  
Challenges include residual entanglement between registers, accumulation of hardware noise in long unrollings, and choices of readout observables \citep{widdows2024near}.  
Despite these limitations, QRNNs provide a natural choice for sequence models and remain a key candidate for quantum NLP.  

\vspace{-1mm}
\subsection{Quantum Convolutional Neural Networks (QCNNs)}
\vspace{-1mm}
QCNNs generalize classical convolutional and pooling operations to quantum circuits, using local entangling unitaries and qubit reduction to extract hierarchical features \citep{cong2019quantum,hur2022quantum}.  
While most prior applications focus on classical data classification, QCNNs have also been adapted for token-level NLP tasks \citep{meichanetzidis2020quantum,widdows2024near}.  
Their parallel structure reduces circuit depth compared to QRNNs, making them a practical choice for NISQ devices, though trade-offs remain between expressivity and hardware feasibility.

\vspace{-1mm}
\section{Method}
\label{sec:method}
\vspace{-1mm}

We introduce hybrid quantum language models (HQLMs) that adapt recurrent and convolutional neural architectures to parameterized quantum circuits (PQCs). %

\vspace{-1mm}
\subsection{Token Embeddings}
\label{sec:embeddings}
\vspace{-1mm}

We adopt $R_y$ angle embeddings as our input encoding. Each token $t \in V$ has a trainable vector $\vtheta_t \in \mathbb{R}^d$, mapped to a separable quantum state
\begin{equation}
    \ket{\psi_t} = \bigotimes_{j=1}^d R_y(\theta_{t,j}) \ket{0}.
\end{equation}
This scheme is shallow, noise-robust, and hardware-friendly, as it avoids entangling gates and allows virtual qubits to be distributed across non-adjacent physical qubits. While richer encodings (e.g.\ amplitude or entangled maps \citep{havlivcek2019supervised,schuld2019quantum, perez2020data,lloyd2020quantum}) exist, they incur greater circuit depth and complexity, leading to increased noise and decoherence. We leave exploration of such embeddings to future work.

\begin{wrapfigure}{r}{0.3\textwidth}
    \centering
    \vspace{-4mm}
    \includegraphics[trim={11mm 5mm 2mm 9mm},clip,width=0.9\linewidth]{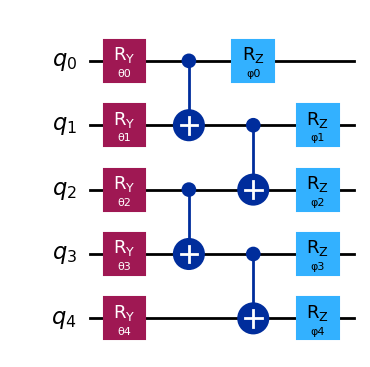}
    \vspace{-2mm}
    \caption{Basic PQC ansatz}%
    \label{fig:pqc_architecture}
    \vspace{-6mm}
\end{wrapfigure}

\vspace{-1mm}
\subsection{PQC Layers as Neural Blocks}
\label{sec:architectures}
\vspace{-1mm}

Following prior QML and QNLP work \citep{sim2019expressibility,schuld2020circuit,bausch2020recurrent,meichanetzidis2020quantum}, we use shallow PQCs as quantum analogues of neural layers. Each layer applies parameterized rotations on each qubit ($R_y$, $R_z$) interleaved with an entangling layer of pairwise CNOTs. This balances expressivity and noise resilience \citep{sim2019expressibility,schuld2021machine}, crucial for NISQ devices where deep entangling networks quickly decohere \citep{preskill2018quantum}. For more expressivity and entangling power, we can stack multiple layers. We use these PQCs as recurrent blocks in QRNNs, convolutional units in QCNNs and prediction heads in both architectures.

\vspace{-1mm}
\subsection{Feature Extraction}
\label{sec:feature-extraction}
\vspace{-1mm}

To map quantum states into features, we consider two strategies:

\textbf{(i) Sampling.} Projective measurement yields bitstrings $\{0,1\}^n$, which can be interpreted as token samples. While conceptually aligned with generative modeling, this method is difficult to train: optimization landscapes become noisy and semantically related tokens may map to distant bitstrings.

\textbf{(ii) Observable estimation.} Measuring expectation values of $Z$ and $ZZ$ operators on hidden registers produces continuous features for a classical linear projection. This yields smoother gradients, better robustness to shot noise, and greater semantic flexibility. 

Although both methods rely on finite sampling in practice, we found observable-based features to be consistently more stable. We therefore use this approach throughout our experiments: we measure $Z$ on all \emph{output} qubits and $ZZ$ on all pairs, yielding a feature vector of size $d + d(d-1)/2$ for $d$ qubits.

\vspace{-1mm}
\subsection{Quantum Recurrent Neural Network (QRNN)}
\label{sec:qrnn}
\vspace{-1mm}

Our QRNN architecture is built around two registers: an embedding register $\mathcal{E}$ and a hidden register $\mathcal{H}$. Each token is encoded into $\mathcal{E}$ with $R_y$ rotations, then transferred to $\mathcal{H}$ through a layer of CNOTs that establish correlations between the new input and the hidden state. The recurrent block $\mathcal{U}_{\text{rec}}$ applies parameterized $R_y$, $R_z$ rotations and entangling gates on $\mathcal{H}$, thereby updating the state across timesteps. The final hidden state is further processed by a separate PQC $\mathcal{U}_{\text{pred}}$ applied on $\mathcal{H}$, and then mapped into $Z$ and $ZZ$ expectation values for the classical projection layer.

\begin{figure}[h]
    \centering
    \includegraphics[trim={11mm 7mm 2mm 6mm},clip,width=0.8\textwidth]{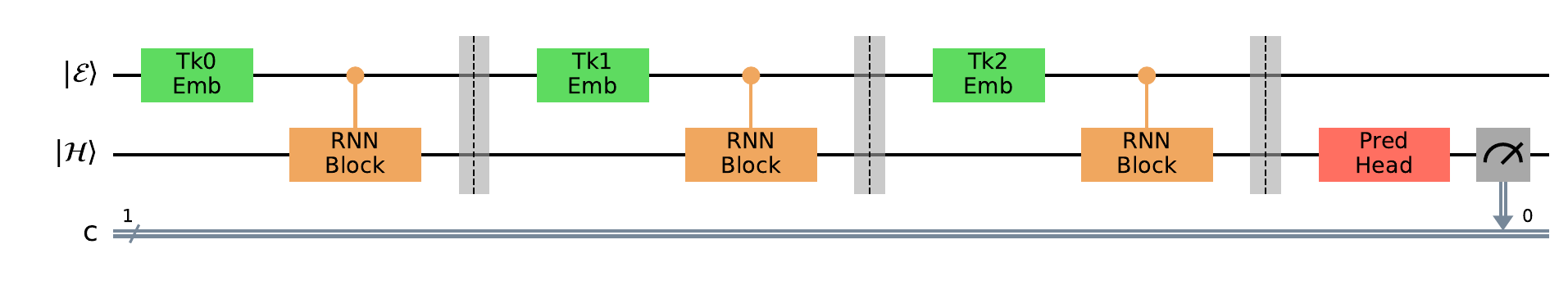}
    \vspace{-2mm}
    \caption{QRNN: tokens are embedded into $\mathcal{E}$, transferred to hidden register $\mathcal{H}$ by CNOTs, updated by recurrent PQC $\mathcal{U}_{\text{rec}}$, and passed through prediction PQC $\mathcal{U}_{\text{pred}}$ for observable-based feature extraction.}
    \label{fig:qrnn_circuit}
    \vspace{-1mm}
\end{figure}

While inspired by prior work on quantum recurrent models \citep{bausch2020recurrent,widdows2024quantum}, our design was adapted specifically to the low-connectivity \emph{heavy-hex} topology of IBM’s Eagle and Heron processors. The placement of CNOT gates and grouping of rotations were chosen to minimize SWAP operations and circuit depth, ensuring better fidelity on real devices. Detailed circuit diagrams, including qubit layouts and hardware mappings, are provided in \cref{app:circuit_details}.  

\vspace{-1mm}
\subsection{Quantum Convolutional Neural Network (QCNN)}
\label{sec:qcnn}
\vspace{-1mm}

\begin{figure}[t]
    \centering
    \includegraphics[trim={11mm 7mm 2mm 6mm},clip,width=0.6\textwidth]{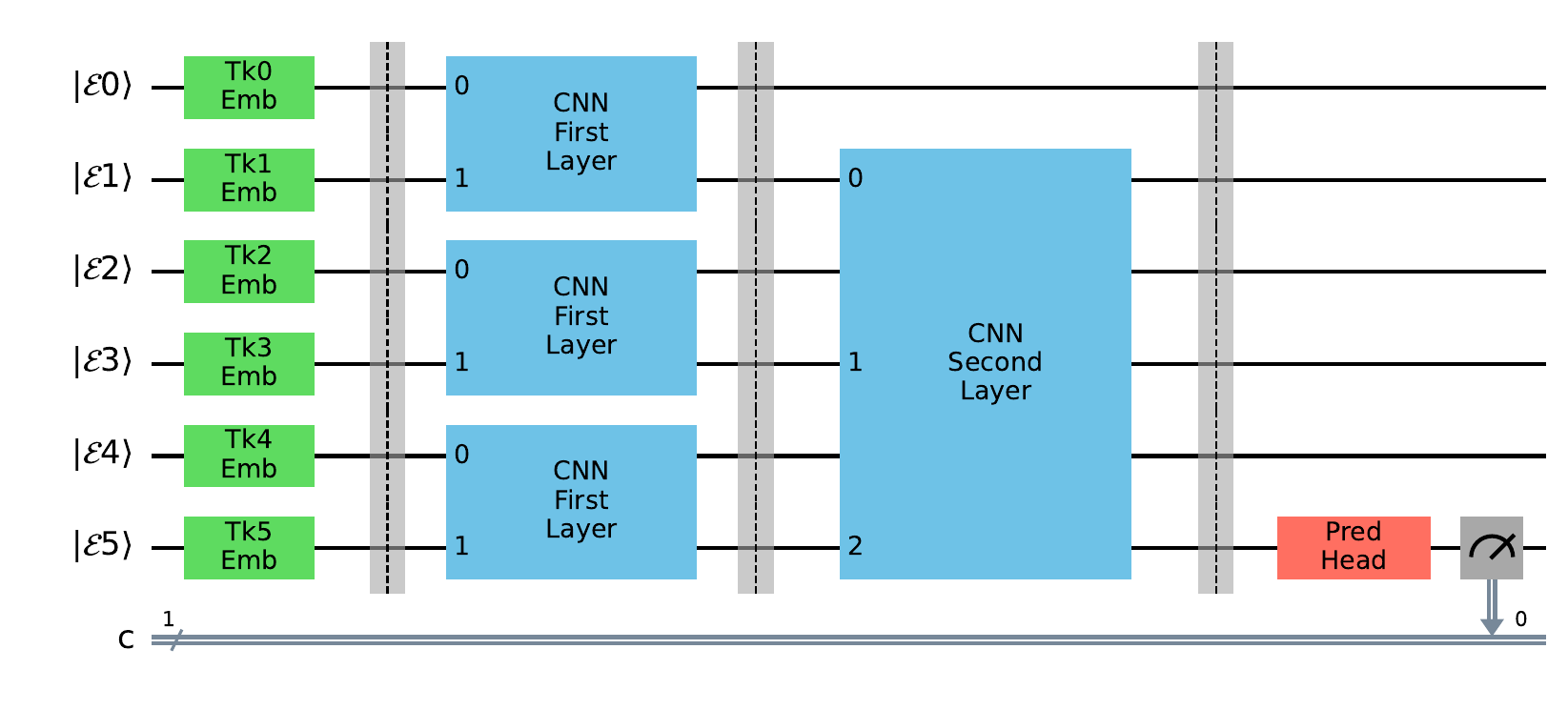}
    \vspace{-2mm}
    \caption{QCNN: tokens are embedded into registers $\mathcal{E}_0...\mathcal{E}_5$, processed by convolutional blocks $\mathcal{U}_{\text{conv}}$, and aggregated via prediction PQC $\mathcal{U}_{\text{pred}}$.}
    \label{fig:qcnn_circuit}
    \vspace{-4mm}
\end{figure}

The QCNN variant follows the principle of local convolutions and pooling \citep{cong2019quantum,hur2022quantum}, but adapted to token-level sequence modeling. Tokens are embedded into parallel registers $\mathcal{E}_0,...,\mathcal{E}_d$, which are grouped into overlapping neighborhoods and processed by convolutional blocks $\mathcal{U}_{\text{conv}}$. Each block processes 2 or 3 adjacent registers and consists of a 2-layered PQC as described in \cref{sec:architectures} to ensure better information flow. Pooling is implemented by only taking 1 register to the next layer, reducing the effective qubit count and yielding hierarchical feature representations. A prediction block $\mathcal{U}_{\text{pred}}$ processes the remaining register, whose state is finally mapped into $Z$ and $ZZ$ observables for classification by the projection head.  

Compared to the sequential QRNN, the QCNN offers shallower depth and greater parallelism, making it attractive for NISQ devices. Our circuit designs were explicitly optimized for IBM hardware by aligning convolutional registers to heavy-hex connectivity and minimizing routing overhead. Full layer diagrams and hardware mapping strategies can be found in \cref{app:circuit_details}.

\vspace{-1mm}
\subsection{Optimization Strategy}
\vspace{-1mm}

Quantum parameters are trained with multi-sample \emph{Simultaneous Perturbation Stochastic Approximation (SPSA)} \citep{spall1998overview}, which estimates gradients by evaluating losses at pairs of symmetric points along random perturbation directions:
\begin{equation}
    \hat{\nabla}_\vtheta L \approx \frac{1}{P} \sum_{p=1}^P \frac{L(\vtheta+\epsilon\vdelta_p)-L(\vtheta-\epsilon\vdelta_p)}{2\epsilon}\,\vdelta_p^{-1}.
\end{equation}
This reduces circuit evaluations compared to parameter-shift or full finite differences while remaining hardware-friendly. The classical projection layer is trained with exact gradients via backpropagation.

\subsection{Expressivity and Non-linearity} \label{sec:expressivity}

A common theoretical concern regarding quantum models is that because unitary evolution is linear in Hilbert space, the resulting models might be limited to linear decision boundaries. However, our hybrid architectures introduce significant non-linearity through two primary mechanisms: the feature map and sequential data re-uploading.

First, the angle embedding strategy ($R_y(\theta)$) maps classical inputs $\theta$ into the amplitudes of the quantum state vector via non-linear trigonometric functions (e.g., $\cos(\theta/2)$). As established by \citet{perez2020data} and \citet{schuld2021effect}, sandwiching these non-linear encoding gates between parameterized unitaries allows a quantum circuit to approximate any continuous function, creating an effect mathematically analogous to activation functions in deep learning.

In our QRNN, this expressivity is naturally amplified by the sequential structure. At each timestep $t$, a new token $x_t$ is encoded and entangled with the persisting hidden state. This repeated injection of data into the evolving quantum state acts as a temporal form of data re-uploading. Consequently, the final hidden state is not a simple linear sum of inputs, but a nested product of non-commuting unitaries dependent on the sequence history. This allows the circuit to model complex, high-frequency correlations in the data.

Finally, the estimator-based readout computes expectation values $f(x) = \langle \psi(x) | \hat{O} | \psi(x) \rangle$, which is a quadratic function of the state amplitudes, providing a final layer of non-linearity before the classical projection head.

\vspace{-1mm}
\section{Experiments}
\label{sec:experiments}
\vspace{-1mm}
We evaluate our proposed hybrid quantum language models on both synthetic datasets and established benchmarks from the quantum natural language processing (QNLP) literature.

\vspace{-1mm}
\subsection{Datasets}
\vspace{-1mm}
\paragraph{Literature benchmarks.}
To enable comparison with prior work, we also evaluate on established QNLP classification datasets \citep{lorenz2023qnlp}:
\begin{itemize}[leftmargin=*, itemsep=0pt, topsep=0pt]
    \item \textbf{MC (Meaning Classification)}: Binary classification with 70 train and 30 test 4-word sentences in two classes: \emph{programming} and \emph{cooking}. We also derive a language modeling version (\textbf{MC-LM}) using the same sentences.
    \item \textbf{RP (Relative Pronoun resolution)}: Binary classification with 74 train and 31 test 4-word sentences, grouped by sentence structure: \emph{X that did Y} vs. \emph{X that Y did}.
\end{itemize}

\paragraph{Synthetic language modeling dataset.}
To evaluate the ability of our models to capture compositional structure in natural language, we also generate a small-scale \textbf{Toy Sentence Language Modeling (TS-LM)} dataset.  
The vocabulary contains 24 unique words, grouped into categories such as \textit{subjects}, \textit{verbs}, \textit{adjectives}, \textit{objects}, \textit{prepositions}, and \textit{locations}.  
Sentences are sampled from a context-free grammar ensuring basic syntactic consistency: \texttt{subject--verb--[adjective]--object--[preposition--location]}.  
For example, a sentence may look like: ``\texttt{man sees small dog on table}''.  
The dataset consists of 200 training and 50 test sentences, with sentence lengths between 3 and 6 words.  
This controlled setup balances variability with tractability for small-scale quantum models.  
The dataset and generation code are available in the supplementary material.

\subsection{Experimental Setup}
\paragraph{Models and Baselines.}
We evaluate:
\begin{itemize}[leftmargin=*, itemsep=0pt, topsep=0pt]
    \item \textbf{Hybrid QLMs (ours)}: QRNN and QCNN architectures (\cref{sec:method}), trained end-to-end with SPSA for quantum parameters and backpropagation for the classical projection head. 
    \item \textbf{Classical baselines}: FFNN, RNN, LSTM, CNN, and Transformer models with comparable parameter counts. 
\end{itemize}

\paragraph{Training Setup.}
All models are trained with Adam.  
Gradients for the quantum parameters are estimated using multi-sample SPSA, while the classical projection head is trained with exact gradients. We use Qiskit \citep{qiskit} for quantum circuit simulation and execution on IBM hardware \citep{ibm2025}, and PyTorch \citep{paszke2019pytorch} for training and evaluating classical models.
Most experiments are performed on simulators; for QRNN and QCNN, we additionally evaluate trained models on real IBM hardware (Eagle/Heron processors). For MC and TS-LM, we also train directly on hardware.  
Embedding registers use \texttt{emb\_size=3} qubits.  
We tune learning rate, batch size, and epochs per task. SPSA uses population size $p=8$ and perturbation scale $\sigma=0.05$.  
Full hyperparameters are given in {\cref{app:hyperparams}}. Detailed information about circuit complexity, number of trainable parameters, and per-epoch training costs for each architecture is provided in \cref{app:training_details}.

\paragraph{Evaluation Protocol.}
For \textbf{language modeling} tasks, we report train/test perplexity (Tr PPL / Ts PPL) and the average probability assigned to the correct next token in the test set (Acc.).  
For \textbf{binary classification} tasks, we report test accuracy.  

\subsection{Main Results}
\Cref{tab:main_results} summarizes the performance of classical and quantum models.  
Classical baselines achieve strong results across tasks, with RNNs, LSTMs, and Transformers performing best overall.  

For quantum models, both QRNN and QCNN achieve accuracies and perplexities comparable to their classical counterparts in simulation. On MC-LM and TS-LM, quantum models reach test perplexities in the same range as classical networks, showing that relatively shallow PQC-based architectures can capture sequential patterns in small-scale language modeling. On the MC classification task, QRNN and QCNN match the perfect accuracy of classical models, while on RP they achieve slightly lower but still competitive accuracies (80.6\% and 83.9\% vs. 90.3\% for RNN). Note however that the RP test set of 31 samples contains 20 words not seen during training, and 4 of the 31 samples cannot be inferred by any model from the train set alone due to inherent ambiguity. Thus, the maximum theoretical accuracy without taking into acount random guesses is 87.1\%, and the QRNN and QCNN achieve scores close to this limit.

Hardware runs highlight the gap between ideal simulation and current devices. Models trained on simulators but evaluated on hardware (\textbf{Real Hardware Eval}) show moderate accuracy/perplexity degradation due to noise which decreases with more circuit samples and better hardware: for example the 5.83 test perplexity of QCNN on MC-LM was obtained with 10k shots on a IBM Heron processor, while lowering the shot count or running on previous generation Eagle processors yields significantly higher perplexities. Fully hardware-trained models (\textbf{Real Hardware Train + Eval}) perform worse still, reflecting harder optimization due to noisy sampling. Notably, QRNNs are somewhat more robust than QCNNs on hardware, likely due to their lower qubit counts and fewer parameters.

Overall, these results indicate that (i) Hybrid quantum language models can match the performance of small classical models on toy NLP benchmarks in simulation. (ii) Noise remains the primary bottleneck for real devices, and (iii) With the development of larger quantum devices, we can move beyond very simple QNLP models such as DisCoCat, which were limited to just a handful of qubits. Our results show that more expressive architectures like QRNNs and QCNNs can be successfully trained on today's hardware. However, careful adaptation of circuit design to the hardware topology (\cref{sec:method}) remains crucial to achieve robust performance.

\begin{table}[t]
\centering
\caption{Performance of classical and quantum language models across multiple tasks. Columns MC and RP report binary classification accuracy (\%), while MC-LM and TS-LM denote two next-token prediction tasks with training/test perplexity and average next token prediction accuracy. For QRNN and QCNN, we report: (i) \textbf{Simulator} results with noiseless statevector simulation; (ii) \textbf{Real Hardware Eval}, where the simulator-trained parameters are executed on quantum processors; and (iii) \textbf{Real Hardware Train + Eval}, where both training and evaluation are carried out directly on quantum hardware.}
\label{tab:main_results}
\vspace{-0.5mm}
\resizebox{1.0\linewidth}{!}{
\begin{tabular}{@{}cclcccccccc@{}}
\toprule
\multirow{2.5}{*}{\makecell{}}          & \multirow{2.5}{*}{\makecell{Training \\ Setup}}             &   \multirow{2.5}{*}{\makecell{Model}}       & \multicolumn{2}{c}{Test Acc. [\%]} & \multicolumn{3}{c}{MC-LM} & \multicolumn{3}{c}{TS-LM} \\ 
\cmidrule(r){4-5} \cmidrule(lr){6-8} \cmidrule(l){9-11}
&  &     & MC              & RP              & Tr PPL  & Ts PPL  & Acc. [\%]   & Tr PPL  & Ts PPL  & Acc. [\%]   \\
\midrule
\multirow{6.5}{*}{\rotatebox[origin=c]{90}{Classical}}  
& \multicolumn{2}{c}{Random Baseline} & 50.0                & 50.0                & 19.0    & 19.0    & 5.26  & 24.0    & 24.0    & 4.17  \\
\cmidrule(lr){2-11}
& \multirow{5}{*}{\makecell{Autograd \\ Backprop}}             
  & FFNN     & 100                 & 93.5                & 3.79    & 6.30    & 20.0  & 3.38    & 3.51    & 31.0  \\
& & RNN      & 100                 & 90.3                & 4.12    & 5.44    & 20.0  & 3.52    & 3.75    & 28.7  \\
& & LSTM     & 100                 & 96.8                & 4.01    & 5.79    & 19.0  & 3.82    & 3.89    & 27.4  \\
& & CNN      & 100                 & 80.6                & 3.80    & 5.07    & 22.0  & 3.73    & 3.65    & 30.8  \\
& & Transf   & 100                 & 83.9                & 3.81    & 4.81    & 23.0  & 3.39    & 3.47    & 32.0  \\
\midrule
\multirow{9}{*}{\rotatebox[origin=c]{90}{Quantum}}   
& \multirow{3}{*}{\makecell{Simulator}}           
  & \textbf{QRNN}     & 100                 & 80.6                & 4.64    & 4.84    & 22.6  & 3.66    & 3.47    & 31.6  \\
& & \textbf{QCNN}     & 100                 & 83.9                & 5.10    & 5.69    & 19.0  & 3.96    & 3.76    & 28.9  \\
& & DisCoCat & 79.8                & 72.3                & /       & /       & /     & /       & /       & /     \\
\cmidrule(l){2-11}
& \multirow{2}{*}{\makecell{Real Hardware \\ Eval}}    
  & \textbf{QRNN}     & 100                 & 74.2                & /       & 4.86    & 22.4  & /       & 3.86    & 28.4  \\
& & \textbf{QCNN}     & 100                 & 77.4                & /       & 5.83    & 18.4  & /       & 4.43    & 25.0  \\
\cmidrule(l){2-11}
& \multirow{3}{*}{\makecell{Real Hardware \\ Train + Eval}}    
  & \textbf{QRNN}     & 100                 & /                   & /       & /       & /     & 4.60    & 4.82    & 24.8  \\
& & \textbf{QCNN}     & 100                 & /                   & /       & /       & /     & 8.82    & 8.65    & 12.4  \\
& & DisCoCat & 83.3                & 67.7                & /       & /       & /     & /       & /       & /     \\
\bottomrule
\end{tabular}}
\end{table}

\clearpage
\subsection{Ablation Studies}

\begin{wraptable}{r}{0.5\textwidth}
\vspace{-4mm}
\centering
\caption{Effect of training randomness. We report the mean and standard deviation of test perplexity or accuracy over 5 independent runs with different random seeds for each of the considered tasks on both QRNN and QCNN architectures.}
\label{tab:randomness}
\resizebox{0.99\linewidth}{!}{
\begin{threeparttable}
\begin{tabular}{llcc}
\toprule
Task        & Model  & Avg Score & Stdev \\ 
\midrule
\multirow{2}{*}{MC (Acc)}    
& QRNN   & 100       & 0.0   \\
& QCNN*  & 85.3      & 19.1  \\
\cmidrule(){1-4}
\multirow{2}{*}{RP (Acc)}    
& QRNN   & 69.6      & 7.1   \\
& QCNN   & 78.7      & 3.7   \\
\cmidrule(){1-4}
\multirow{2}{*}{MC-LM (PPL)} 
& QRNN   & 5.00      & 0.19  \\
& QCNN** & 6.28      & 0.62  \\
\cmidrule(){1-4}
\multirow{2}{*}{TS-LM (PPL)} 
& QRNN   & 4.12      & 0.62  \\
& QCNN   & 4.28      & 0.62  \\ 
\bottomrule
\end{tabular}
\begin{tablenotes}[flushleft]
    \setlength\leftmargini{0pt}
    \footnotesize
    \item * 3 of 5 runs have scores of 100\%, others fail to learn 
    \item ** 1 of 5 runs had a PPL of 13.3 so we consider it an outlier
\end{tablenotes}
\end{threeparttable}
}
\vspace{-4mm}
\end{wraptable}

\paragraph{Training Randomness} In \cref{tab:randomness}, we study the effect of training randomness by reporting the mean and standard deviation of test perplexity or accuracy over 5 independent runs with different random seeds for each of the considered tasks on both QRNN and QCNN architectures. We observe that both models exhibit rather high variance across runs, indicating sensitivity to initialization and stochasticity in the training process. The effect is more pronounced for QCNN, which has higher qubit and quantum parameter counts, leading to a higher dimensionality of the optimization landscape and therefore more challenges in finding optimal solutions, including very weak gradient signal in barren plateaus. This suggests that further work is needed to improve training stability, potentially through better initialization schemes, regularization techniques, or more robust optimization methods.

\begin{figure}[t]
    \centering
    \begin{subfigure}{0.48\linewidth}
        \centering
        \includegraphics[width=.88\linewidth]{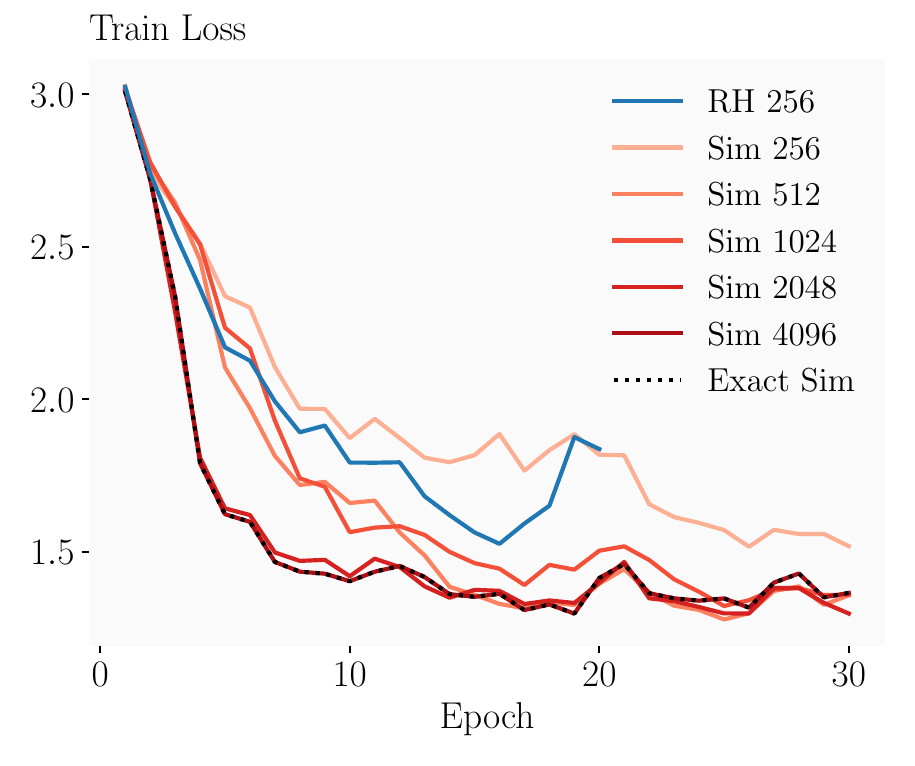}
        \vspace{-3mm}
        \caption{Train Loss evolution for different shot counts used for estimation in both Simulator and Real Hardware.}
        \label{fig:shots_plot}
    \end{subfigure}
    \hfill
    \begin{subfigure}{0.48\linewidth}
        \centering
        \includegraphics[width=.88\linewidth]{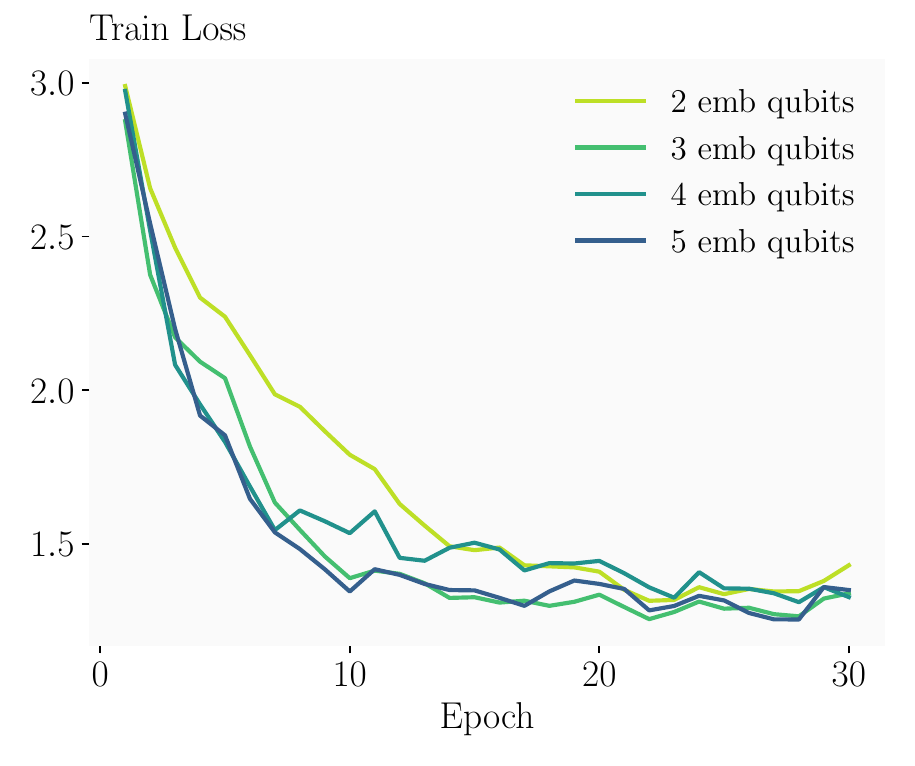}
        \vspace{-3mm}
        \caption{Train Loss evolution for different embedding sizes (number of qubits).}
        \label{fig:embedding_size_plot}
    \end{subfigure}
    \vspace{-1.5mm}
    \caption{Train Loss evolution for different hyperparameters when training QRNN models on TS-LM.}
    \label{fig:ablation}
    \vspace{-5.5mm}
\end{figure}

\vspace{-1mm}
\paragraph{Number of shots} 
In \cref{fig:shots_plot}, we show how the number of shots used for expectation estimation affects QRNN training. Increasing shots generally speeds up training by providing more accurate loss estimates, with diminishing returns beyond 4096 shots. Even 256 shots allow effective learning, demonstrating robustness to shot noise. Training on real hardware with 256 shots yields comparable performance to simulations, highlighting that near-term devices can provide useful gradient information despite noise.

\paragraph{Embedding Size} 
In \cref{fig:embedding_size_plot}, we illustrate the effect of embedding size (number of qubits per embedding register) on QRNN training. Larger embeddings reduce training loss by enabling more expressive representations, but gains saturate beyond a certain point, indicating an optimal qubit count. Note that the total qubit count includes both embedding and hidden registers: for example, 2 embedding qubits → 4 total, 5 → 12 total, and 8 → 19 total. While larger embeddings improve expressivity, they also increase the number of quantum parameters and the complexity of the optimization landscape, making training more challenging. On simulators, larger embeddings increase computational cost exponentially (20 s/epoch, 65 s/epoch, 1150 s/epoch for the three examples), whereas on real hardware timing remains roughly constant ($\approx$1000 s/epoch) due to communication overheads. This highlights the trade-off between expressivity, trainability, and practical efficiency when choosing embedding size.

\section{Limitations} \label{sec:discussion}

While our results demonstrate the feasibility of generative QNLP on near-term devices, several limitations remain.
First, \textbf{hardware noise} creates a significant gap between simulation and execution. QRNNs, despite their parameter efficiency, suffer from decoherence in deeper sequential unrollings, whereas QCNNs require larger qubit counts that introduce crosstalk errors.
Second, \textbf{optimization stability} varies significantly by architecture; QCNNs exhibit higher variance across initialization seeds (\cref{tab:randomness}), suggesting that barren plateaus or local minima remain a challenge for wider variational circuits trained via SPSA.
Third, \textbf{vocabulary scalability} presents a fundamental bottleneck for generative quantum models. Using observable-based readout requires scaling the number of measurements or the classical projection layer linearly with vocabulary size, which restricts current implementations to small synthetic lexicons.
Finally, regarding \textbf{expressivity}, while our use of sequential data encoding naturally induces non-linearity via data re-uploading, shallow realizations of these circuits may essentially function as kernel methods. Scaling to deeper, more expressive circuits to capture complex semantic dependencies requires further advances in error mitigation to survive the noise floor of current NISQ processors.

\section{Conclusions} \label{sec:conclusions}

We presented the first empirical demonstration of hybrid quantum sequence models trained and evaluated end-to-end on real quantum hardware for generative tasks. By adapting QRNN and QCNN architectures to the heavy-hex topology of IBM processors and utilizing a scalable SPSA-based training workflow, we established a performance baseline for generative QNLP in the NISQ era.

Key engineering insights from this study include:
(i) \textbf{Estimator-based readout} significantly outperforms bitstring sampling for gradient-based training, providing the smoothness required for convergence on noisy devices;
(ii) \textbf{Hardware-aware circuit mapping} is critical: naive logical mapping fails due to SWAP overhead, whereas our topology-aligned designs enable effective learning; and
(iii) There exists a clear \textbf{trade-off between circuit depth and width}: QRNNs offer qubit efficiency but suffer from depth-dependent noise accumulation, while QCNNs offer shallower depth at the cost of higher connectivity requirements.

Our work validates that hybrid quantum architectures can learn syntactic structures despite hardware imperfections, provided that the training stack is carefully optimized for the device constraints. Future work will focus on overcoming the vocabulary bottleneck through hierarchical quantum embeddings and exploring spatially multiplexed attention mechanisms to further parallelize execution.

\section*{Reproducibility Statement}

We release the complete code used for our experiments at 
\href{https://github.com/stefanrzv2000/Hybrid-QLM}{github.com/stefanrzv2000/Hybrid-QLM}. 
A detailed description of the experimental setup and hyperparameters is provided in \cref{app:exp_details}.

\ificlrfinal
	\section*{Acknowledgements}

We acknowledge the use of IBM Quantum services for this work. The views expressed are those of the authors, and do not reflect the official policy or position of IBM or the IBM Quantum team.

\else
	\ificlrpreprint
		\section*{Acknowledgements}

We acknowledge the use of IBM Quantum services for this work. The views expressed are those of the authors, and do not reflect the official policy or position of IBM or the IBM Quantum team.

	\fi
\fi
\message{^^JLASTBODYPAGE \thepage^^J}

\bibliography{references}
\bibliographystyle{iclr2026_conference}

\ifincludeappendixx
	\clearpage
	\appendix
	\onecolumn
	
\section{LLM usage disclaimer}

This paper was prepared using the following Large Language Models (LLMs):  GPT 4.1 and GPT-5 (OpenAI). The authors have reviewed and edited the content produced by these models, and take full responsibility for the final content of the publication. The LLMs were used to assist with drafting and editing text, improving grammar and style, and suggesting rephrasings. No scientific claims or results were generated by the LLMs; all technical content, experiments, and conclusions are the original work of the authors. The use of LLMs was limited to non-technical writing tasks to enhance clarity and readability. Additionally, the LLMs were used for searching literature for relevant papers and resources. The authors acknowledge the potential limitations and biases of LLMs, and have carefully verified all information in the final manuscript. Any errors or inaccuracies are solely the responsibility of the authors.

\section{Notation}
\label{app:notation}

\subsection{Quantum Registers and Parameter Notation}
\label{sec:qc-registers}

In our hybrid quantum language models, we explicitly distinguish between different quantum registers, each with its own purpose and number of qubits. Specifically, we consider:

\begin{itemize}
    \item \textbf{Embedding register} $\mathcal{E}$ with $d_e$ qubits, which encodes the input token embeddings.
    \item \textbf{Hidden register} $\mathcal{H}$ with $d_h$ qubits, which stores recurrent or latent states.
    \item \textbf{Output or prediction register} $\mathcal{O}$ with $d_o$ qubits, optionally used for feature extraction or measurement.
\end{itemize}

Each register has its own Hilbert space, and the total system is represented as
\begin{equation}
\mathcal{H}_\text{total} = \mathcal{H}_\mathcal{E} \otimes \mathcal{H}_\mathcal{H} \otimes \mathcal{H}_\mathcal{O} \cong \mathbb{C}^{2^{d_e + d_h + d_o}}.
\end{equation}

Individual qubits within a register are indexed by lowercase letters, e.g., $q_j \in \mathcal{E}$ or $h_k \in \mathcal{H}$. A tensor-product quantum state can be written as
\begin{equation}
\ket{\psi_\text{in}} = \bigotimes_{j=1}^{d_e} R_y(\theta_{j}) \ket{0}_{q_j} \otimes \bigotimes_{k=1}^{d_h} \ket{0}_{h_k}.
\end{equation}

To simplify notation, we often write the all-zero state of $d$ qubits as $\ket{\vzero_d}$, or simply $\ket{\vzero}$ when the dimension is clear from context.

\paragraph{Parameter vectors.} 
Token embeddings are represented by trainable vectors $\vtheta_v \in \mathbb{R}^{d_e}$ for each token $v \in V$, which are mapped to the embedding register via $R_y$ rotations. Hidden registers and entangling layers are parametrized separately, e.g., $\vphi \in \mathbb{R}^{n_h}$. For convenience, all trainable quantum parameters can be concatenated into a master vector $\vTheta = [\vtheta, \vphi]$ when describing gradient updates or optimization procedures.

\paragraph{Observables and measurements.} 
When measuring quantum states, we explicitly specify the register of interest. For example, the $Z$ observable over the hidden register is denoted
\begin{equation}
\hat{Z}_{\mathcal{H}} = \sum_{j \in \mathcal{H}} Z_j,
\end{equation}
which is used to extract features for the classical output layer. This notation clarifies which registers contribute to the model output, particularly in hybrid architectures with multiple quantum sub-registers.

\subsection{Quantum Gates and Circuit Operations}
\label{sec:qc-gates}

Quantum circuits manipulate qubits via unitary gates. Single-qubit rotations, particularly $R_y$ gates, are used to encode token embeddings:
\begin{equation}
R_y(\theta_j) \ket{0}_{q_j} = \cos(\theta_j/2)\ket{0}_{q_j} + \sin(\theta_j/2)\ket{1}_{q_j}, \quad q_j \in \mathcal{E}.
\end{equation}

Multi-qubit interactions are introduced through entangling gates, such as controlled-NOT (CNOT) operations:
\begin{equation}
\text{CNOT}_{c,t} \ket{q_c q_t} = 
\begin{cases}
\ket{q_c q_t \oplus 1}, & \text{if } q_c = 1,\\
\ket{q_c q_t}, & \text{otherwise},
\end{cases}
\end{equation}
where $c$ and $t$ denote control and target qubits, respectively. Entangling gates are primarily applied between embedding and hidden registers, or within the hidden register, to capture correlations across tokens.

We denote the full parametrized circuit acting on registers $\mathcal{E}$ and $\mathcal{H}$ as
\begin{equation}
U(\vTheta) = \prod_{l=1}^{L} U_l(\vTheta_l),
\end{equation}
where each layer $U_l$ can contain both single-qubit rotations and entangling operations, and $\vTheta_l \subset \vTheta$ represents the parameters in that layer. The resulting quantum state is then
\begin{equation}
\ket{\psi(\vTheta)} = U(\vTheta) \ket{\vzero},
\end{equation}
with $\ket{\vzero}$ the all-zero state of the entire system, as defined previously.

\paragraph{Observables.} 
Measurement operators are associated with specific registers to extract features for the classical output layer. For example, for the hidden register $\mathcal{H}$, $Z$ and $ZZ$ operators are used to compute expectation values:
\begin{equation}
f(\vTheta) = \langle \psi(\vTheta) | \hat{O}_\mathcal{H} | \psi(\vTheta) \rangle,
\end{equation}
where $\hat{O}_\mathcal{H}$ denotes the collection of observables applied to $\mathcal{H}$. These expectation values serve as inputs to classical layers, providing a hybrid quantum-classical representation.

\section{Experimental Details}
\label{app:exp_details}

\subsection{Hyperparameters and Training Details}
\label{app:hyperparams}
Detailed hyperparameters for all experiments are summarized in \cref{tab:hyperparams}. We tune learning rate, batch size, and number of epochs per task. SPSA uses population size $p=8$ and perturbation scale $\sigma=0.05$.

\begin{table}[h]
    \centering
    \caption{Hyperparameters for all experiments.}
    \label{tab:hyperparams}
    \resizebox{0.8\linewidth}{!}{
    \begin{tabular}{lcccccc}
        \toprule
        Task & Model & Seq Len & CNN kernels & Learning Rate & Batch Size & Epochs \\
        \midrule
        MC & QRNN & 4 & /   & 0.1 & 10 & 20 \\
        MC & QCNN & 6 & 3,3 & 0.1 & 10 & 20 \\
        RP & QRNN & 4 & /   & 0.1 & 10 & 40 \\
        RP & QCNN & 4 & 2,2 & 0.1 & 10 & 40 \\
        MC-LM & QRNN & 4 & /   & 0.1 & 16 & 40 \\
        MC-LM & QCNN & 6 & 3,3 & 0.1 & 16 & 40 \\
        TS-LM & QRNN & 6 & /   & 0.1 & 32 & 30 \\
        TS-LM & QCNN & 6 & 3,3 & 0.1 & 32 & 30 \\
        \bottomrule
    \end{tabular}}
\end{table}

\subsection{Quantum Processors Used}
\label{app:quantum_hardware}
We utilize IBM's quantum processors for our experiments, specifically the Eagle and Heron architectures. A list of all the processors used, along with their key specifications, is provided in \cref{tab:quantum_hardware}.

\begin{table}[h]
    \centering
    \caption{IBM Quantum processors used in experiments.}
    \label{tab:quantum_hardware}
    \resizebox{0.9\linewidth}{!}{
    \begin{tabular}{llcccc}
        \toprule
        Processor & Architecture & Qubits & Topology & Median T1 (µs) & Median 2Q Error \\
        \midrule
        \texttt{ibm\_pittsburgh} & Heron r3 & 156 & Heavy-hex & 316 & $1.5\cdot10^{-3}$ \\
        \texttt{ibm\_kingston} & Heron r2 & 156 & Heavy-hex & 261 & $1.8\cdot10^{-3}$ \\
        \texttt{ibm\_fez} & Heron r2 & 156 & Heavy-hex & 145 & $2.7\cdot10^{-3}$ \\
        \texttt{ibm\_aachen} & Heron r2 & 156 & Heavy-hex & 205 & $1.7\cdot10^{-3}$ \\
        \texttt{ibm\_strasbourg} & Eagle r3 & 127 & Heavy-hex & 295 & $8.7\cdot10^{-3}$ \\
        \texttt{ibm\_brussels} & Eagle r3 & 127 & Heavy-hex & 245 & $7.9\cdot10^{-3}$ \\
        \bottomrule
    \end{tabular}}
\end{table}

\subsection{Training Details}
\label{app:training_details}

Detailed information about circuit complexity, number of trainable parameters, and per-epoch training costs for each architecture is provided in \cref{tab:training_details}.

\begin{table}[h]
\centering
\caption{Model and training details for all architectures on the TS-LM task. Time per epoch is measured on an AMD Ryzen AI 9HX with 32GB RAM for classical models and simulators, and IBM Eagle/Heron processor for quantum models (in parentheses, the actual quantum processing time excluding overhead). For quantum models, we report two embedding sizes (E=3,10 for QRNN, E=3,6 for QCNN). For QRNN E=10 we have 17 hidden qubits, but we limit the $ZZ$ observables to pairs of adjacent qubits only, reducing the final feature size to 19.}
\label{tab:training_details}
\resizebox{0.98\linewidth}{!}{
\begin{tabular}{@{}lllccccc@{}}
\toprule
\multirow{2.5}{*}{Type} & \multirow{2.5}{*}{Model} & \multirow{2.5}{*}{\makecell{Params \\ Total (Q+C)}}         & \multicolumn{4}{c}{Quantum Circ}           & \multirow{2.5}{*}{\makecell{Time per epoch \\ clock time (Q usage)}}       \\ 
\cmidrule(lr){4-7}
&    &     & Qubits & Total gates & 2Q gates & 2Q depth &  \\
\midrule
\multirow{5}{*}{Classical}
& FFNN     & 316 (0+316)    &        &             &          &          & \textless{}1s        \\
& RNN      & 256 (0+256)    &        &             &          &          & \textless{}1s        \\
& LSTM     & 376 (0+376)    &        &             &          &          & \textless{}1s        \\
& CNN      & 304 (0+304)    &        &             &          &          & \textless{}1s        \\
& Transf   & 384 (0+384)    &        &             &          &          & \textless{}1s        \\
\midrule
\multirow{4}{*}{\makecell{Quantum \\ Simulator}}      
& QRNN (E=3)  & 258 (90+168)   & 6      & 106         & 34       & 22       & 20s                  \\
& QCNN (E=3)  & 316 (148+168)  & 19     & 181         & 48       & 12       & 110s                 \\
& QRNN (E=10) & 1158 (342+816) & 27     & 1966        & 188      & 22       & \textgreater{}2h     \\
& QCNN (E=6)  & 1044 (340+704) & 37     & 1213        & 102      & 12       & OOM                  \\
\midrule
\multirow{4}{*}{\makecell{Quantum \\ Hardware}}  
& QRNN (E=3)  & 258 (90+168)   & 6      & 370         & 34       & 22       & 16min (7min)         \\
& QCNN (E=3)  & 316 (148+168)  & 19     & 577         & 48       & 12       & 35min (7min)         \\
& QRNN (E=10) & 1158 (342+816) & 27     & 1966        & 188      & 22       & 40min (8min)         \\
& QCNN (E=6)  & 1044 (340+704) & 37     & 1213        & 102      & 12       & 80min (8min)         \\ 
\bottomrule
\end{tabular}}
\end{table}

\subsection{Circuit Details}
\label{app:circuit_details}

\cref{fig:qrnn_detailed} and \cref{fig:qcnn_detailed} show detailed circuit diagrams for QRNN and QCNN architectures, respectively. Example qubit layouts on IBM Heron processor for different embedding sizes are shown in \cref{fig:rnn_hardware} and \cref{fig:cnn_hardware}.

\clearpage
\begin{figure}[t]
    \centering
    \includegraphics[width=0.99\textwidth]{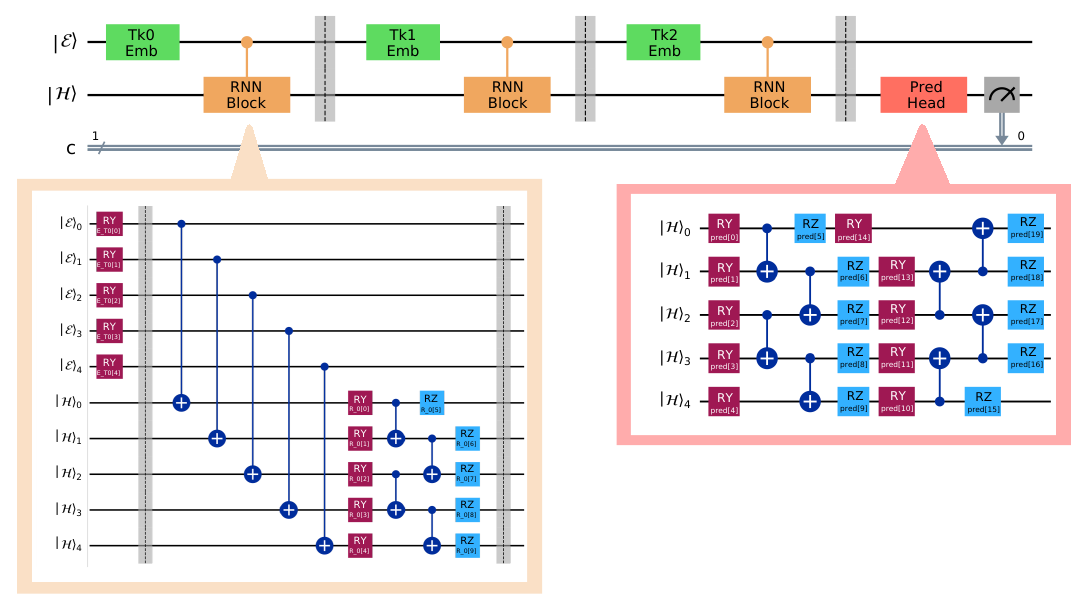}
    \vspace{-2mm}
    \caption{QRNN: tokens are embedded into $E$, transferred to hidden register $H$ by CNOTs, updated by recurrent PQC $\mathcal{U}_{\text{rec}}$, and passed through prediction PQC $\mathcal{U}_{\text{pred}}$ for observable-based feature extraction.}
    \label{fig:qrnn_detailed}
    \vspace{-4mm}
\end{figure}

\begin{figure}[t]
    \centering
    \begin{subfigure}{0.49\linewidth}
        \centering
        \includegraphics[width=.99\linewidth]{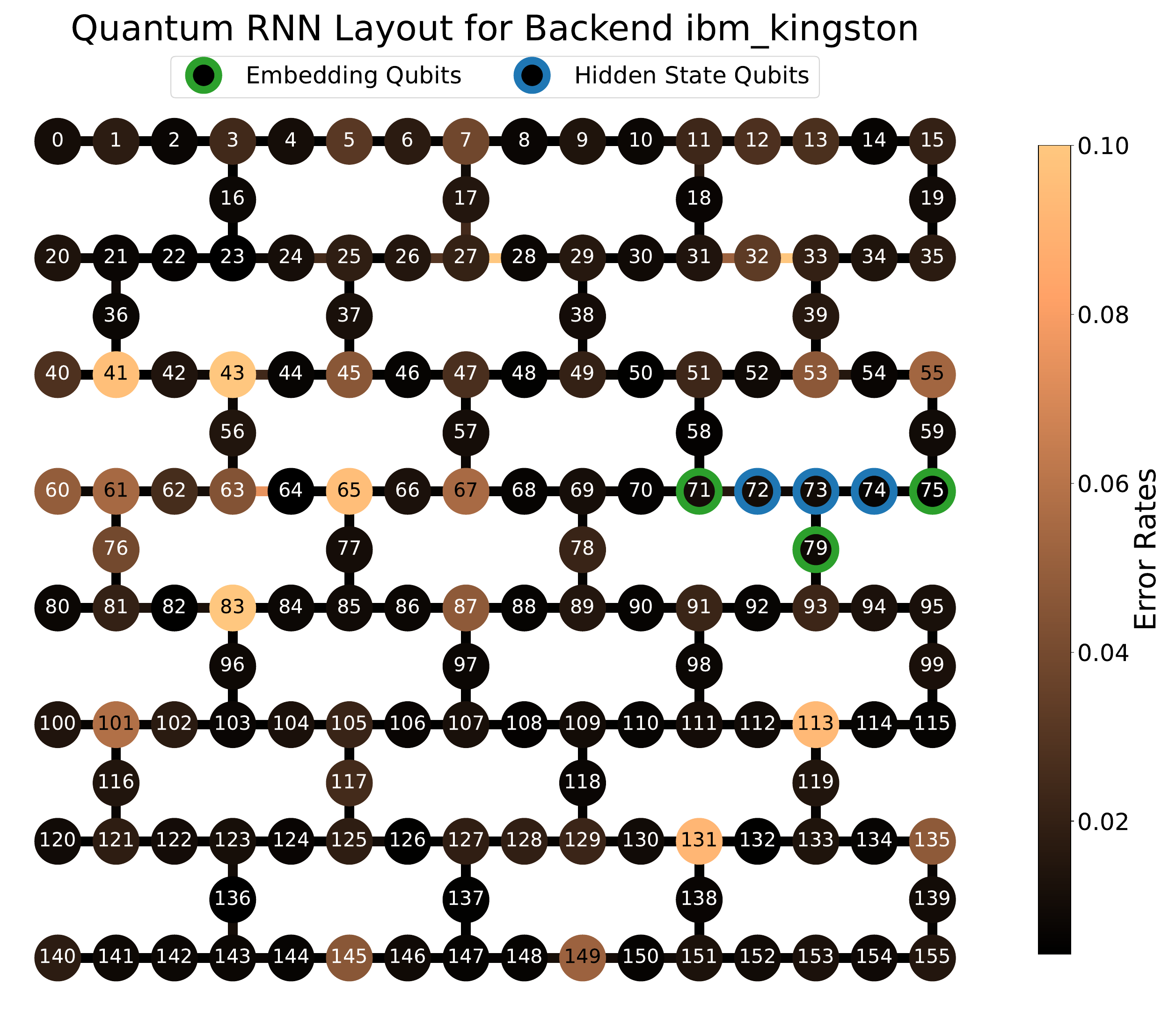}
        \vspace{-3mm}
        \caption{3 embedding + 3 hidden qubits}
        \label{fig:rnn3hardware}
    \end{subfigure}
    \hfill
    \begin{subfigure}{0.49\linewidth}
        \centering
        \includegraphics[width=.99\linewidth]{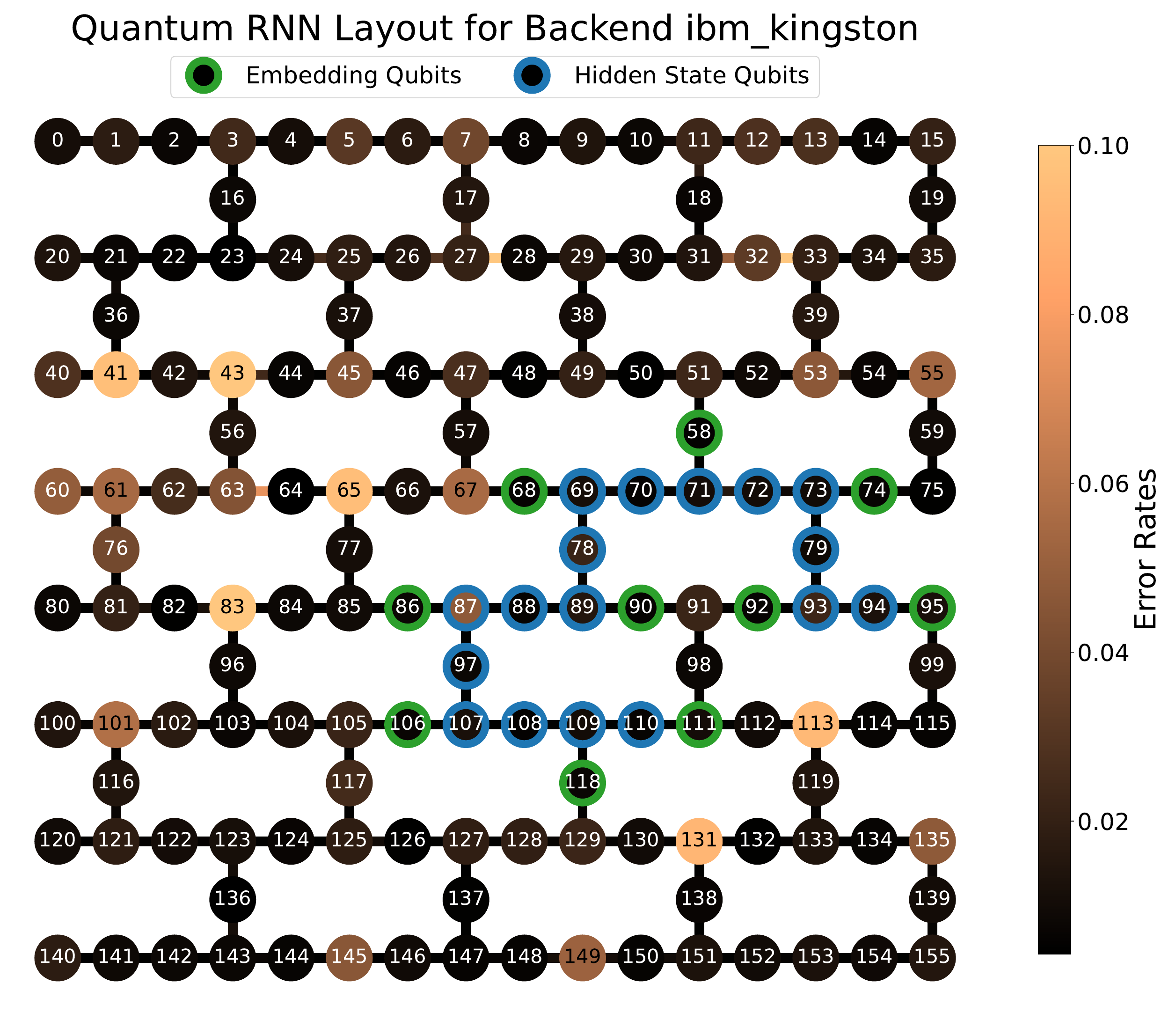}
        \vspace{-3mm}
        \caption{10 embedding + 17 hidden qubits}
        \label{fig:rnn10hardware}
    \end{subfigure}
    \vspace{-1mm}
    \caption{Example qubit layout on IBM Heron processor for QRNN with (a) 3-qubit embedding and (b) 10-qubit embedding. The heavy-hex connectivity is highlighted, along with error rates for single and 2 qubit gates. \textbf{\textcolor{green}{Green}} qubits are used for embedding register $\mathcal{E}$ and are not connected. \textbf{\textcolor{blue}{Blue}} qubits are used for hidden register $\mathcal{H}$ and need to be connected. Some of the hidden qubits are auxiliary and do not correspond to embedding qubits in order to respect the hardware connectivity.}
    \label{fig:rnn_hardware}
    \vspace{-5mm}
\end{figure}

\begin{figure}[t]
    \centering
    \includegraphics[width=0.99\textwidth]{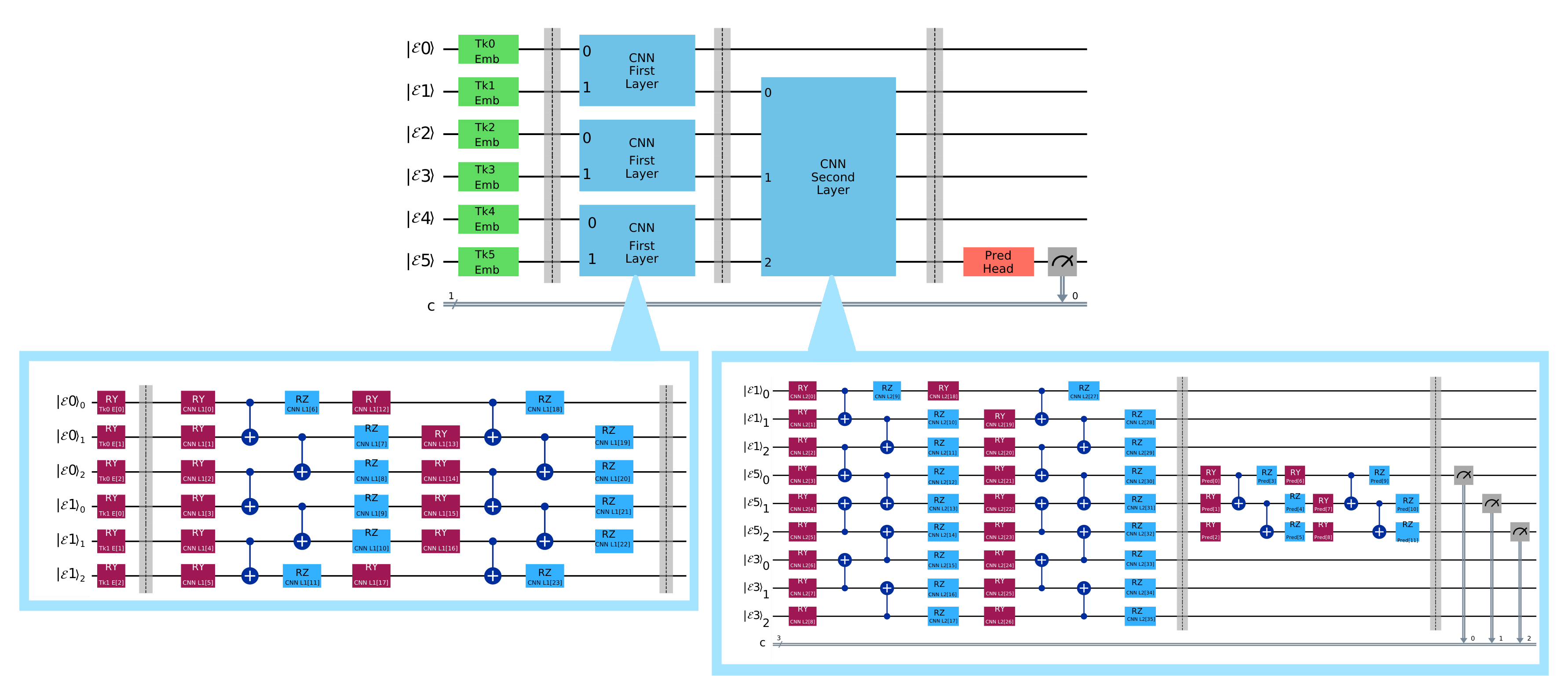}
    \vspace{-2mm}
    \caption{QCNN: tokens are embedded into registers $\mathcal{E}_0...\mathcal{E}_5$ then processed by convolutional blocks $\mathcal{U}_{\text{conv1}}$ and $\mathcal{U}_{\text{conv2}}$, and prediction PQC $\mathcal{U}_{\text{pred}}$.}
    \label{fig:qcnn_detailed}
    \vspace{-4mm}
\end{figure}

\begin{figure}[t]
    \centering
    \begin{subfigure}{0.49\linewidth}
        \centering
        \includegraphics[width=.99\linewidth]{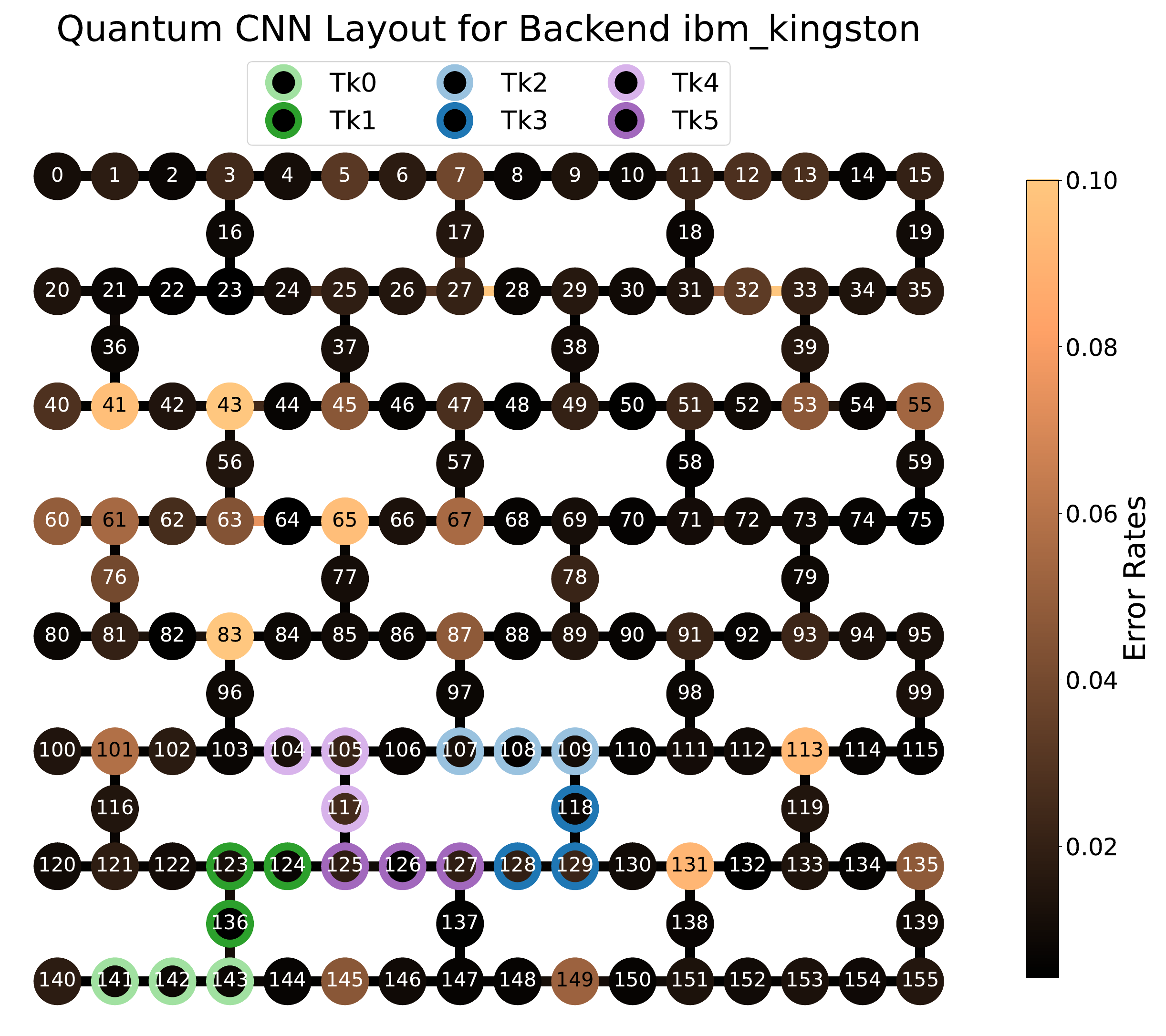}
        \vspace{-3mm}
        \caption{$6\times3$ embedding qubits}
        \label{fig:cnn3hardware}
    \end{subfigure}
    \hfill
    \begin{subfigure}{0.49\linewidth}
        \centering
        \includegraphics[width=.99\linewidth]{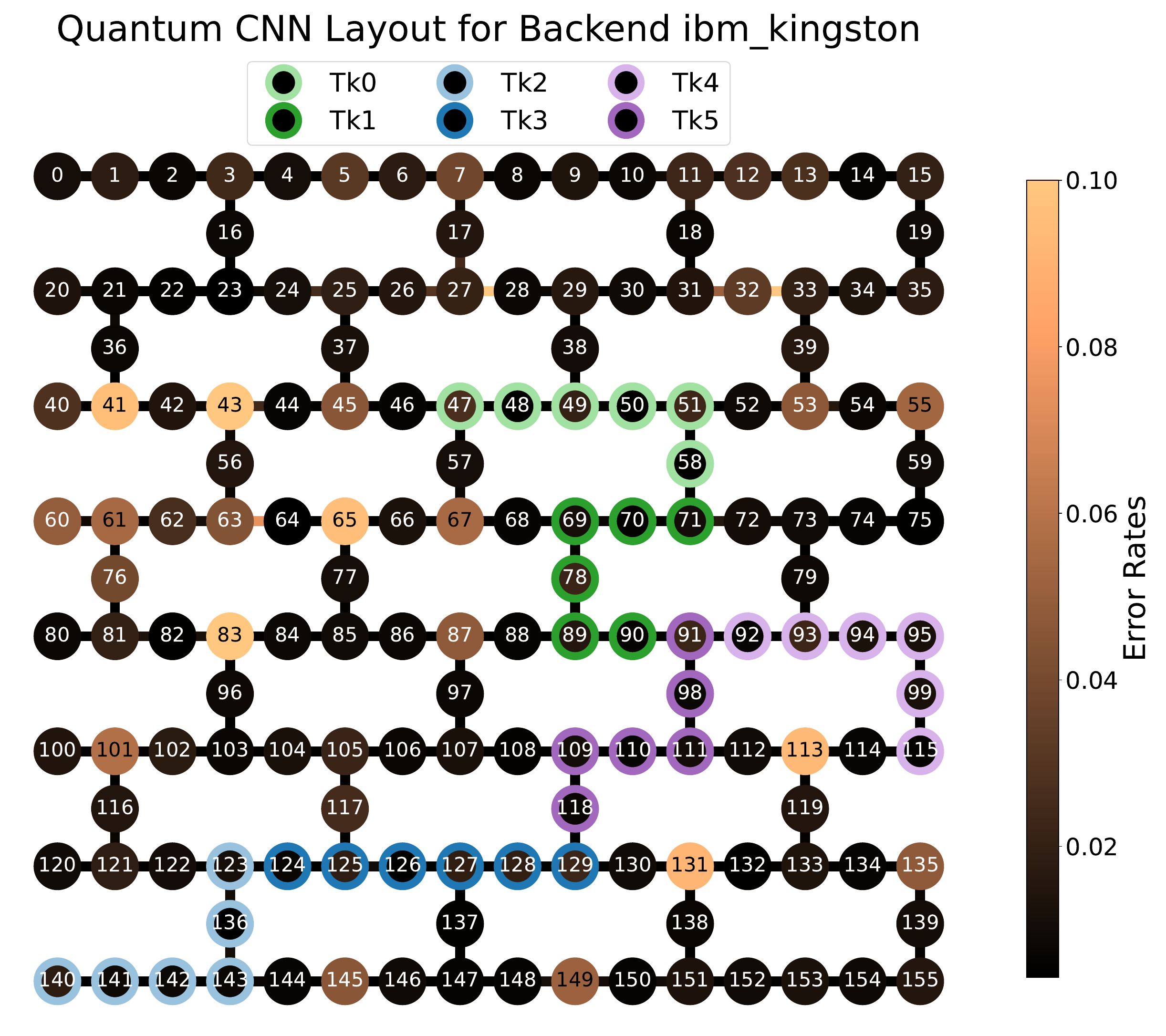}
        \vspace{-3mm}
        \caption{$6\times6$ embedding qubits}
        \label{fig:cnn6hardware}
    \end{subfigure}
    \vspace{-1mm}
    \caption{Example qubit layout on IBM Heron processor for QCNN with (a) $6\times3$ embedding and (b) $6\times6$ embedding. The heavy-hex connectivity is highlighted, along with error rates for single and 2 qubit gates. Each color represents one embedding register $\mathcal{E}_i$, where qubits need to be connected. The two shades of each color represent the connections made by the first convolutional block. The darker shade of each color represents the qubits used in the second convolutional block. \textbf{\textcolor{purple}{Dark Purple}} qubits are used for measurements as prediction register $\mathcal{O}$.}
    \label{fig:cnn_hardware}
    \vspace{-5mm}
\end{figure}

\fi

\end{document}